\documentstyle[12pt,pictex]{article}
\def\square{\vcenter{\vbox{\hrule height.4pt
          \hbox{\vrule width.4pt height8pt
          \kern8pt\vrule width.4pt}\hrule height.4pt}}}

\begin{document}

\centerline{\Large\bf The Free Energy of High Temperature QED}
\centerline{\Large\bf to Order $e^{5}$ From Effective Field
Theory}

\vskip 10mm
\centerline{Jens O. Andersen}
\centerline{\it Institute of Physics}
\centerline{\it University of Oslo}
\centerline{\it P.O. BOX 1048, Blindern}
\centerline{\it N-0316 Oslo, Norway}

\begin{abstract}
{\footnotesize Massless quantum electrodynamics is studied at
high temperature
and zero chemical potential. We compute the Debye screening
mass
to order $e^{4}$ and the free energy to order $e^{5}$}
by an effective field theory
approach, recently developed by Braaten and Nieto.
Our results are in agreement with calculations done in
resummed
perturbation theory. This method makes it possible to
separate
contributions to the free energy from different momentum
scales
(order $T$ and $eT$)
and provides an economical
alternative to computations in the full theory which
involves the
dressing of internal propagators.\\ \\
PACS number(s): 11.10.Wx, 12.20.DS, 12.38.Bx
\end{abstract}
\small
\normalsize
\section{Introduction}
Effective field theories both at zero and finite temperature
have become of increasing interest in recent years [1-3].
Modern developments in renormalization theory have given
meaning
to non-renormalizable quantum
field theories (``effective theories'') and it is no longer
obvious that
renormalizabilty is an essential property of a useful field
theory \cite{lepage}.
The distinction between effective theories and ``fundamental''
(renormalizable)
field theories becomes rather vague. Indeed, the modern view
is to consider every field theory as effective, valid and with
predictive power at a certain scale.

There has been a tremendous progress in perturbative
calculations
of the thermal properties
for quantum field theories at high temperatures since the
pioneering work of Dolan and Jackiw some twenty years ago
 \cite{dolan}. Today, one can essentially distinguish between
two methods for computing high temperature
properties of a quantum field theory. The first method is based
on resummed perturbation theory. Resummation is a reorganization
of the perturbation series in order to be actually perturbative.
Resummation gives rise to effective thermal masses
(and in non-Abelian theories effective vertices, too) that
provide
an infrared
cut off in the loops. The dressed propagators are thus well
behaved
and the use of these in loops yields the contributions
to the free energy that are non-analytic in $g^{2}$, where
$g$ is the coupling constant. The use of resummed perturbation
theory amounts to summing an infinite
set of higher order diagrams.
This resummation scheme has largely been developed by
Braaten and Pisarski \cite{pis} and has been applied to
QCD by a number of authors in connection with the gauge
dependence of the gluon damping rate and the screening mass
(Refs. [7,8] and references therein). The method has also been
used to to calculate the free energy in QED [9,10] and in the
standard model \cite{arnold12}.

The second method is based on dimensional reduction
\cite{lands}.
The crucial observation is that the only contribution to a
correlator that
is not exponentially damped at scales larger than $T$ comes
from the
static bosonic modes. One strategy is therefore to integrate
out the non-static bosonic
modes as well as the fermionic modes to construct an effective
three-dimensional field theory of the zero modes, in which the
masses and couplings are temperature dependent.
This approach has been used in the study of high temperature
QCD and QED \cite{lands}
as well as in investigations
of phase transitions in spontaneously broken gauge
theories [1,13]
(in particular the electro-weak phase transition).

Instead of explicitly integrating over the heavy modes (which
produces
non-local terms beyond one-loop \cite{jacovac3}), one simply
writes down the most general three-dimensional Lagrangian,
that respects the symmetries
of the system. One then computes static correlators in the two
theories and
require that they match. This matching requirement can
actually be taken
as the definition of the effective field theory and provides
the relationship
between the coupling constants in the effective theory and
the underlying theory.
This method has previously been applied to massless $\phi^{4}$
theory
\cite{braaten} and QCD \cite{braaten2}. In the latter
case it does not only provide a convenient
way of calculating the free energy in the high temperature
limit, it does
also solve the infrared catastrophe of Non-Abelian gauge
theories. It is a well known fact that the free energy of
Non-Abelian gauge theories may
be calculated to fifth order in the coupling using resummed
perturbation theory \cite{linde}. However, the method
breaks down at order $g^{6}$ and
this has been interpreted as a sign that the transverse
gluons have a mass of order $g^{2}T$, which cannot be
calculated
in perturbation theory \cite{arnold12}. This magnetic
screening
mass should then provide the necessary infrared cutoff.

In the present paper we would like to apply the method to
QED and
confirm some recent results obtained by calculations in
the full theory
[9,10]. Our work will serve as a nice demonstration of the
efficiency of the effective field theory approach by
allowing one to work with a single scale at a time.

The outline of this work is as follows: In section two we
compute the screening mass to fourth
order and the free energy to fifth order in the coupling
constant in high temperature QED. In section
three we summarize. Throughout the work we use the imaginary
time formalism and the theory is regularized by working in
$4-2\epsilon$
dimensions together with the $\overline{MS}$
regularization scheme.
We use the following shorthand notation
for the sum-integrals that appear below:
\begin{eqnarray}
\int\{dK\}f(K)&\equiv &\mu^{2\epsilon}
T\!\!\!\!\!\!\!\!\!\!\!\!
\sum_{k_{0}=2\pi (n+1/2)T}\int\frac{d^{3-2\epsilon}k}
{(2\pi)^{3-2\epsilon}}f(K),
\end{eqnarray}
where $K^{2}=k_{0}^{2}+k^{2}$.
In the Feynman
diagrams a solid lines denotes a fermion, a wavy line a
photon
and a dashed line a ghost.
\section{QED at High Temperature}
The partition function can  be written as a path integral
\begin{equation}
\label{z1}
{\cal Z}=\int {\cal D}A_{\mu}\,{\cal D}\overline{\psi}
\,{\cal D}
\psi \exp\Big[-\int_{0}^{\beta}d\tau\int d^{3}x\,
{\cal L}\Big ],
\end{equation}
where the Lagrangian of QED reads
\begin{equation}
{\cal L}=\frac{1}{4}F_{\mu\nu}F_{\mu\nu}+\overline{\psi}
\gamma^{\mu}\Big (\partial_{\mu}-ieA_{\mu} \Big )\psi
+{\cal L}_{\mbox{\footnotesize gf}}.
\end{equation}
In this paper we choose to work in Feynman gauge where
$L_{\mbox{\footnotesize gf}}=
\frac{1}{2}(\partial_{\mu}A_{\mu})^{2}$.
In the effective theory the partition function reads
\begin{equation}
\label{z2}
{\cal Z}=e^{-f(\Lambda )V}\int {\cal D}A_{i}\,{\cal D}
A_{0}\exp
\Big[-\int d^{3}x\,{\cal L}_{\mbox{\footnotesize eff}}\Big].
\end{equation}
Here ${\cal L}_{\mbox{\footnotesize eff}}$ is the effective three-dimensional
Lagrangian to which
we shall return shortly.
The prefactor $f(\Lambda )$ can be interpreted as the
coefficient
of the unit
operator in the effective field theory. It depends on the
ultraviolet cutoff $\Lambda$ (which is introduced to regularize
the effective three-dimensional theory) in order to cancel the
$\Lambda$ dependence
in the path
integral in Eq. (\ref{z2}) \cite{braaten}.
The effective three-dimensional theory consists of a gauge
field
and a real
massive self interacting scalar field in the adjoint
representation
of the gauge group:
\begin{equation}
{\cal L}=\frac{1}{4}F_{ij}F_{ij}+\frac{1}{2}
(\partial_{i}\rho)^{2}
+\frac{1}{2}m^{2}(\Lambda )\rho^{2}+\lambda (\Lambda)\rho^{4}
+{\cal L}_{\mbox{\footnotesize gf}}+\delta {\cal L}.
\end{equation}
At this point a few comments are in order. Firstly, we remark
that
the scalar field $\rho$ is identified with the time component
$A_{0}$ of the gauge
field in the full theory, up to normalizations. This implies
that
the field $A_{0}$ does not interact with
the gauge field in the effective theory, since QED is an
Abelian
gauge theory.
The fact that $A_{0}$  may develop a thermal mass is simply
a consequence of the lack of Lorentz invariance at non-zero
temperature. Secondly, it can easily be
demonstrated from the matching of four-point functions that
$\lambda (\Lambda)$ is
of order $e^{4}$ and that the term $\lambda
(\Lambda)\rho^{4}$ does not
contribute to the Debye mass or the free energy to the order
we are calculating (This interaction contributes at
order $e^{5}$
to the screening mass and at order $e^{6}$ to the
free energy).

\begin{figure}[b]
\underline{FIGURE CAPTIONS:}
\caption{Self-energy correction in the effective theory.}
\label{qed2a}
\caption{One-loop self-energy correction.}
\label{qed5}
\caption{One-loop diagrams.}
\label{qed1}
\caption{Two-loop diagram.}
\label{qed2}
\caption{Three-loop diagrams.}
\label{qed3}
\caption{One-loop diagram with mass insertion.}
\label{qed4}
\end{figure}
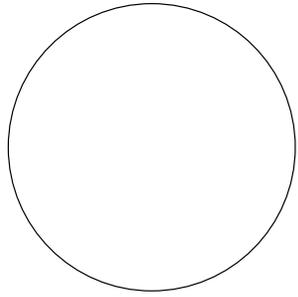
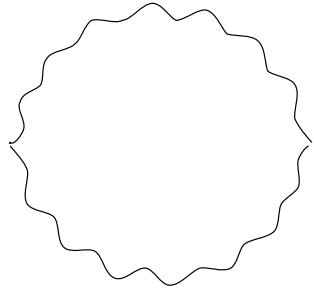
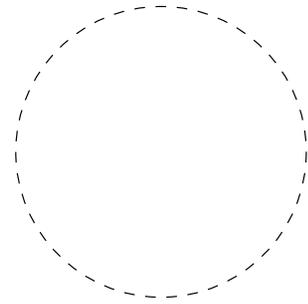
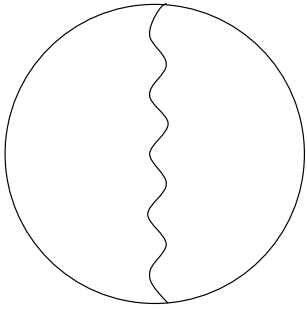
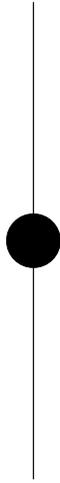
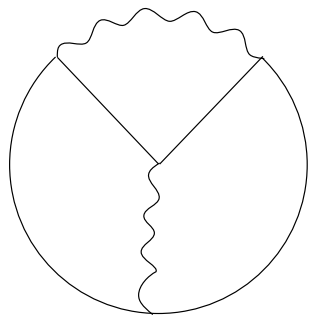
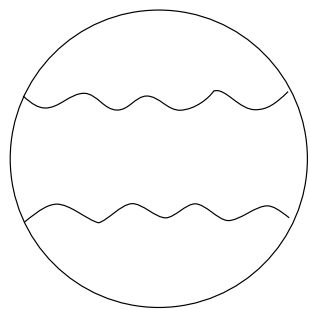
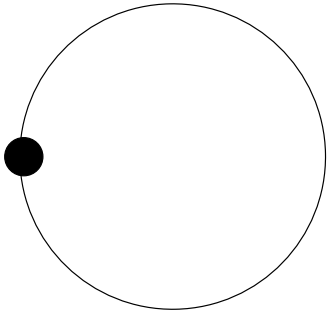
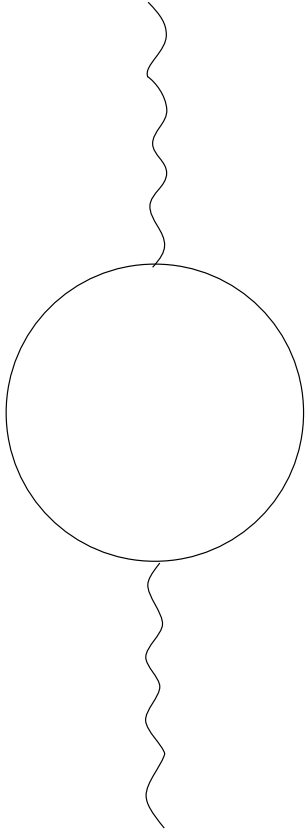

Furthermore, $\delta {\cal L}$ represents all local
terms which
respect
the symmetries of the theory such as gauge invariance. This
includes
renormalizable terms, such as $g(\Lambda )\rho^{6}$, as well
as
non-renormalizable ones.\\ \\
{\it The mass parameter.}$\,\,$
In order to obtain the free energy to order $e^{5}$, we
need to
 compute the
short distance coefficient $m^{2}(\Lambda )$ to order
$e^{4}$.
The $e^{3}$ and $e^{5}$
contributions to $\ln {\cal Z}$ is namely given by a one-loop
calculation
in the effective theory (see below).
The simplest way to determine $m^{2}(\Lambda )$ in the
effective
theory is by
demanding that the screening mass in the two theories match.
The screening mass of the particles is defined as the location
of the
pole of the propagator for spacelike momentum \cite{braaten}:
\begin{equation}
p^{2}+\Pi_{00} (0,{\bf p})=0,\hspace{1cm}p^{2}=-m^{2}_{s}.
\end{equation}
This definition is gauge fixing independent, which one can
prove on an algebraic level \cite{kobes}. This is in
contrast with the one found in the book by Kapusta
\cite{kapusta} and which commonly used in the
literature$\!\!$
\footnote{In quantum electrodynamics the latter {\it is}
gauge fixing independent, but it is {\it not} renormalization
group invariant. In non-Abelian theories it is also gauge
dependent.
This is to be expected since it is an off-shell
quantity \cite{anton1}.}:
\begin{equation}
m^{2}_{s}=\Pi_{00} (p_{0}=0,{\bf p}\rightarrow 0).
\end{equation}
Normally, as in the present case, the two definitions are
equivalent
to leading order in the coupling constant. Beyond leading
order
they do not coincide.
The requirement above implies that
\begin{equation}
p^{2}+m^{2}(\Lambda )+\Pi_{\mbox{\footnotesize eff}}(p,\Lambda )=0,
\hspace{1cm}p^{2}=-m^{2}_{s},
\end{equation}
where $\Pi_{\mbox{\footnotesize eff}}(p,\Lambda )$
is the self-energy of $\rho$
in the effective
theory. We shall do the matching in strict perturbation
theory. Generally, this means that the expression for the
screening mass differs
from the one obtained in resummed perturbation
theory, which
correctly incorporates the effects of electrostatic
screening. The solution $m_{s}^{2}$ is of order $e^{2}$,
which implies that one can expand $\Pi (p^{2})\equiv
\Pi_{00}(0,p)$
in a Taylor series around $p^{2}=0$. To determine the
screening mass (in strict perturbation theory) to
order $e^{4}$, we must calculate
$\Pi (p^{2})$ to one loop order and $\Pi '(0)$
to two loop order, and the screening mass is then given by
\begin{equation}
\label{scrm}
m_{s}^{2}\approx\Pi (0)\Big [1-\Pi '(0)\Big].
\end{equation}
The symbol $\approx$ indicates that Eq. (\ref{scrm})
only holds in
strict perturbation theory.
Now, strict perturbation theory in the effective theory
means
that the mass term should be treated as a perturbation.
The corresponding contribution to
$\Pi_{\mbox{\footnotesize eff}}$ is
shown in
Fig.~\ref{qed2a}, where the blob indicates the mass insertion.
The matching relation simply becomes
$m^{2}(\Lambda )\approx m_{s}^{2}$.\\ \\
The self-energy to one-loop order in the full theory reads
\begin{eqnarray}\nonumber
\label{scr1}
\Pi (p^{2}) &=&e^{2}\int \{dK\}Tr(\gamma_{0}\frac{1}
{K\!\!\!\!/}\gamma_{0}\frac{1}{K\!\!\!\!/+P\!\!\!\!/})\\
&=&-e^{2}\int \{dK\}\frac{8}{K^{2}}+e^{2}\frac{4}{3}p^{2}
\int \{dK\}\frac{1}{K^{4}}+{\cal O}(p^{4}).
\end{eqnarray}
The corresponding Feynman diagram is shown in
Fig.~\ref{qed5}.
The sum-integrals in Eq. (\ref{scr1}) are standard and can be
found in e.g. Ref.~\cite{arnold1}.
Carrying out the renormalization of the wave function in the
usual
way, one finds:
\begin{equation}
\label{1loop}
\Pi (p^{2})=\frac{e^{2}T^{2}}{3}+\frac{p^{2}}{12\pi^{2}}
(2\gamma_{E} -1+2\ln \frac{\Lambda }{4\pi T}).
\end{equation}
Here $\Lambda$ is the scale introduced by dimensional
regularization.
The two-loop expression for the self energy at zero external
momentum can be found
either by a direct computation of the two-loop graphs or by
applying the
formula (see Ref.~\cite{kapusta})
\begin{equation}
\Pi (0)=-e^{2}\frac{\partial^{2}P}{\partial\mu^{2}},
\end{equation}
where $P$ is the pressure and $\mu$ is the chemical
potential.
This requires
the calculation of the free energy to two-loop order
including
the chemical
potential. This can be carried out by contour integration
\cite{kapusta}.
Using the
result $-T^{2}e^{4}/8\pi^{2}$ for the $e^{4}$ contribution,
we find
\begin{equation}
\Pi (p^{2})=\frac{e^{2}T^{2}}{3}+\frac{e^{2}p^{2}}{12\pi^{2}}
(2\gamma_{E} -1+2\ln \frac{\Lambda}{4\pi T})
-\frac{e^{4}}{8\pi^{2}}.
\end{equation}
The mass parameter to order $e^{4}$ then becomes:
\begin{equation}
\label{mass}
m^{2}(\Lambda)=T^{2}\Big [\frac{e^{2}}{3}
-\frac{e^{4}}{36\pi^{2}}(2\gamma_{E} -1+2
\ln \frac{\Lambda}{4\pi T}) -\frac{T^{2}
e^{4}}{8\pi^{2}}\Big ].
\end{equation}
By using the renormalization
group equation for the coupling constant,
\begin{equation}
\mu\frac{de^{2}}{d\mu}=\frac{e^{4}}{6\pi^{2}}+{\cal O}
(e^{6}),
\end{equation}
one can easily demonstrate that Eq. (\ref{mass}) is
independent of $\Lambda$.
Thus, up to corrections of order $e^{6}$,
we can replace $\Lambda$ by an arbitrary
renormalization scale $\mu$. \\ \\
{\it The coupling constant.}$\,\,$
We would also like to make a few remarks about the gauge
coupling
$e_{3}$ in the effective theory.
Using the relation between the gauge fields in the two
theories
\begin{eqnarray}
A_{i}^{3}&=&\sqrt{T}A_{i},
\end{eqnarray}
one finds that $e^{2}_{3}=Te^{2}$ to leading order in
$e^{2}$. There is no dependence of $\Lambda$ at this order.
Beyond leading order one must compute
and match four point correlation functions
\cite{braaten}. \\ \\
{\it The coefficient of the unit operator.}$\,\,$
We shall now compute $f(\Lambda )$ to order $e^{4}$ in strict
perturbation
theory. A strict perturbative expansion in the coupling
constant is normally
plagued by severe infrared divergences, which are caused by
long-range forces.
Physically, these forces are screened at the length scale
$1/(eT)$ and can
be taken into account only by reorganizing perturbation theory
which amounts
to summing an infinite set of loops. Nevertheless, the strict
perturbation theory
can be used to compute the short distance coefficient
$f(\Lambda )$
and we
shall do so by matching calculations of $\ln {\cal Z}$ in the
full theory and in the effective theory. From eqs. (\ref{z1})
and (\ref{z2}),
we see that the matching condition reads
\begin{equation}
\ln {\cal Z}=-f(\Lambda )V+\ln
{\cal Z}_{\mbox{\footnotesize eff}}.
\end{equation}
The calculation of $\ln {\cal Z}$ in the full theory
involves one-loop, two-loop and three-loop diagrams.
The graphs are displayed in Figs.~\ref{qed1}$-$\ref{qed3}.
The diagrams can be computed by the methods developed in Refs.
[9,19] and details may be found there. We find:
\begin{eqnarray}
\label{unit}\nonumber
\frac{T\ln {\cal Z}}{V}&\approx&\frac{\pi^{2}T^{4}}{9}\Big
[\frac{11}{20}-\frac{5e^{2}}{32\pi^{2}}+
\frac{e^{4}}{256\pi^{4}}\Big ( -\frac{20}{3}
\ln (\frac{\Lambda}{4\pi T}) +\frac{8}{3}
\frac{\zeta '(-3)}
{\zeta (-3)}-\frac{16}{3}\frac{\zeta '(-1)}{\zeta (-1)}
\Big.\Big. \\
&&\Big.\Big.-4\gamma_{E} -\frac{319}{12} +\frac{208}{5}
\ln 2\Big )\Big ].
\end{eqnarray}
Here $\approx$ again indicates that Eq.~(\ref{unit})
only holds in
strict perturbation theory, and
$\Lambda$ is the momentum scale which is introduced by
dimensional regularization. $\Lambda$ may again be traded
for
an arbitrary $\mu$ by RG-arguments.

We now turn to the effective theory.
The mass parameter is viewed as a perturbation in the
effective theory.
$\ln {\cal Z}_{\mbox{\footnotesize eff}}$
is then given
by one-loop contributions from the gauge field and scalar
field
(Figs.~\ref{qed1}$-$\ref{qed3}), plus an
additional one-loop diagram with a mass insertion
(which is indicated by a blob in Fig.~\ref{qed4}).
The computation is
rather simple since one-loop contributions involving massless
fields vanish
identically in dimensional regularization
(see Eq. (\ref{1l}) below)
and so does $\ln {\cal Z}_{\mbox{\footnotesize eff}}$.
The matching condition
therefore turns out to be
\begin{equation}
\frac{T\ln {\cal Z}}{V}\approx-f(\Lambda )T,
\end{equation}
and $f(\Lambda)$ is given by the right hand side of Eq.
(\ref{unit}).
The function $F=f(\Lambda) T$ can be viewed as the
contribution to the free
energy from the short distance scale $1/T$.
With the comments after Eq. (\ref{unit}) in mind, it is
clear that
$f(\Lambda )$ has no dependence of $\Lambda$
at the order we are calculating.

Now that we have determined the short distance coefficients
$m^{2}(\Lambda )$ and $f(\Lambda)$ in the effective
theory, we can calculate the screening mass as
well as the free energy in QED to order $e^{4}$ and
$e^{5}$, respectively. \\ \\
{\it The screening mass.}$\,\,$
In order to calculate the {\it physical}
screening mass, one must include the mass parameter
$m^{2}(\Lambda )$ in the free
part of the Lagrangian. Furthermore,
since the term $\lambda (\Lambda)$ goes like $e^{4}$,
the $\phi^{4}$ term
does not contribute to the screening mass to order
$e^{4}$, as we mentioned
above. This implies that
the physical screening mass\footnote{It is
also equal to the screening
mass obtained in the strict perturbation theory to order
$e^{4}$. This
is due to the fact that there is no $e^{3}$ term,
as explained above. At order $e^{5}$ and higher they
differ, caused by the dressing of bosonic propagators in
the
two-loop diagrams.} is equal to the
short distance coefficient $m^{2}(\Lambda )$
to order e$^{4}$:
\begin{equation}
m_{s}^{2}=T^{2}\Big [\frac{e^{2}}{3}-\frac{e^{4}}
{36\pi^{2}}(2\gamma_{E} -1+2\ln \frac{\bar{\mu}}
{4\pi T}) -\frac{T^{2}e^{4}}{8\pi^{2}}\Big ].
\end{equation}
It is easily checked that the result is RG-invariant
as required. Furthermore,
our result agrees with the calculation of Blaizot
{\it et al} to order
$e^{4}$ \cite{parw1}. Note also that there is no
$e^{3}$ term in the expression
for the screening mass in contrast with both
$\phi^{4}$ theory and SQED.
The reason is that there are no bosonic propagators
in the one-loop self-energy graph in QED and fermions
need no resummation, since their Matsubara frequencies
are never zero.\\ \\
{\it The free energy.}$\,\,$
The calculation of the free energy in the effective
theory is now
straightforward, it is simply a one-loop computation.
However, we must now
include the physical effect of screening, which
amounts to consider the
mass parameter as a term in the free part of the Lagrangian.
Using dimensional regularization (see e.g. Ref.~\cite{ryder})
the one-loop
integrals are perfectly finite after regularization and
independent of the
renormalization scale:
\begin{equation}
\label{1l}
\int \frac{d^{3}k}{(2\pi)^{3}}\ln (k^{2}+M^{2})=
-\frac{1}{6\pi}M^{3}.
\end{equation}
The contribution from the gauge field vanishes. Using
the expression for the mass of the scalar field and
expanding it
in powers of $e$ yields the following contribution to the
free energy:
\begin{equation}
\label{logeff}
\ln {\cal Z}_{\mbox{\footnotesize eff}}=
T^{4}\Big [\frac{e^{3}}{36\sqrt{3}\,\pi}
-\frac{e^{5}}{576\sqrt{3}\pi^{3}}\Big(4\ln\frac{\bar{\mu}}
{4\pi T}+4\gamma_{E}+8\ln 2+9\Big)\Big ].
\end{equation}
This term takes into account the effects from long distance
scales of order
$1/(eT)$. Note that it is non-analytic in $e^{2}$. Furthermore,
in resummed perturbation
theory the $e^{3}$ and the $e^{5}$ terms would arise from the
dressing of
the photon propagator in the two and three-loop diagrams,
respectively.
Putting the results from Eq.s (\ref{unit}) and
(\ref{logeff})
together, one obtains
\begin{eqnarray}\nonumber
\label{res}
\ln{\cal Z}&=&T^{4}\Big[\frac{11\pi^{2}}{180}
-\frac{5e^{2}}{248}
+\frac{e^{3}}{36\sqrt{3}\,\pi}+\frac{e^{4}}{2304\pi^{2}}
\Big ( -\frac{20}{3}\ln (\frac{\bar{\mu}}{4\pi T}) +
\frac{8}{3}\frac{\zeta '(-3)}{\zeta (-3)}\Big.\Big.
-\frac{16}{3}\frac{\zeta '(-1)}{\zeta (-1)}\\\nonumber
&&\Big.\Big.-4\gamma_{E} -\frac{319}{12} +\frac{208}{5}
\ln 2\Big )\Big.\Big.-\frac{e^{5}}{576\sqrt{3}\pi^{3}}
\Big(4\ln\frac{\bar{\mu}}{4\pi T}+4\gamma_{E}+8\ln 2+
9\Big )\Big ] .
\end{eqnarray}
This is the main result of the present paper and it is in
agreement with the computation of Zhai
and Kastening \cite{kast},
and Parwani \cite{parw2}
who use resummed perturbation theory. Eq. (\ref{res})
is correct as well as
renormalization group invariant up to
corrections of order $e^{6}\ln e$. The latter property may
easily be checked by using
the one-loop $\beta $-function in QED.

Finally, we would like to comment upon a computational
difference between
QED and QCD. In QCD the computation of the free energy
involves
the construction of {\it two} effective field theories, which
reflects the fact that
there are contributions from three different momentum scales
($T$, $gT$ and
$g^{2}T$, where $g$ is the gauge coupling) \cite{braaten2}.
The first effective field
theory, called electrostatic QCD (EQCD), consists of the
magnetostatic $A_{i}^{a}$ field and the electrostatic field
$A_{0}^{a}$. The unit operator
$f_{\mbox{\footnotesize QCD}}$ as well as the
mass parameter in EQCD
is then determined by the matching procedure
and $f_{\mbox{\footnotesize QCD}}$ gives the contribution
to the free energy
from the short
distance scale $1/T$.
The second effective field theory is called magnetostatic
QCD (MQCD) and consists
simply of the self interacting magnetostatic gauge field
$A_{i}^{a}$.
Again the unit operator of this effective theory,
$f_{\mbox{\footnotesize EQCD}}$,
can be determined
and yields the contribution to the free energy from the
distance scale
$1/(gT)$. Now, the
perturbative expansion in MQCD is plagued with infrared
divergences, implying
that the functional integral can only be calculated
non-perturbatively, e.g.
by putting MQCD on a lattice. Using lattice simulations
the path integral
may be computed, so that one obtains the contribution
to the free energy
from the scale $1/(g^{2}T)$. One can, of course, also
construct a second
effective field theory in QED, but it is completely
unnecessary.
The
restriction to the magnetostatic zero modes simply yields
a free photon field
theory in three dimensions. This in turn implies that
there are no contribution
from momentum scales $e^{2}T$ in QED. The same remark
applies to scalar electrodynamics.
\section{Summary}
In this work we have computed the free energy at high
temperature
in massless quantum electrodynamics to order $e^{5}$
using an effective
field theory approach. This approach makes it possible
to separate
contributions
which come from different length scales ($1/T$ and $1/(eT)$).
 Furthermore, this method involves
considerably less effort in obtaining the free energy to
a given order
in perturbation theory than computations in the full four
dimensional theory.
\\ \\
The author would like to thank F. Ravndal for comments
and suggestions, and H. Haugerud for carefully reading the
manuscript.

\newpage
\pagebreak

\end{document}